\documentclass[twocolumn,superscriptaddress]{revtex4-1}
\usepackage{bbm}

\def\s{\sigma}

\newcommand{\tw}{t_\mathrm{w}}
\newcommand{\Leff}{L_\mathrm{eff}}
\usepackage[T1]{fontenc}
\usepackage{graphicx}
\usepackage{amsmath}
\usepackage{amssymb}
\usepackage{hyperref}
\makeatletter
\newcommand*{\balancecolsandclearpage}{%
  \close@column@grid
  \cleardoublepage
  \twocolumngrid
}
\makeatother

\begin{document}
\title{A statics-dynamics equivalence through the fluctuation-dissipation ratio provides a window into the spin-glass phase from nonequilibrium measurements}

\author{M.~Baity-Jesi}\affiliation{Institut de Physique Th\'eorique, DRF, CEA Saclay, F-91191 Gif-sur-Yvette Cedex, France}
\author{E.~Calore}\affiliation{Dipartimento di Fisica e Scienze della Terra, Universit\`a di Ferrara e INFN, Sezione di Ferrara, Ferrara, Italy}
\author{A.~Cruz}\affiliation{Departamento  de F\'\i{}sica Te\'orica, Universidad de Zaragoza, 50009 Zaragoza, Spain}\affiliation{Instituto de Biocomputaci\'on y F\'{\i}sica de Sistemas Complejos (BIFI), 50009 Zaragoza, Spain}
\author{L.A.~Fernandez}\affiliation{Departamento  de F\'\i{}sica Te\'orica I, Universidad Complutense, 28040 Madrid, Spain}\affiliation{Instituto de Biocomputaci\'on y F\'{\i}sica de Sistemas Complejos (BIFI), 50009 Zaragoza, Spain}
\author{J.M.~Gil-Narvion}\affiliation{Instituto de Biocomputaci\'on y F\'{\i}sica de Sistemas Complejos (BIFI), 50009 Zaragoza, Spain}
\author{A.~Gordillo-Guerrero}\affiliation{Departamento de  Ingenier\'{\i}a El\'ectrica, Electr\'onica y Autom\'atica, U. de Extremadura, 10071, C\'aceres, Spain}\affiliation{Instituto de Biocomputaci\'on y F\'{\i}sica de Sistemas Complejos (BIFI), 50009 Zaragoza, Spain}
\author{D.~I\~niguez}\affiliation{Instituto de Biocomputaci\'on y F\'{\i}sica de Sistemas Complejos (BIFI), 50009 Zaragoza, Spain}\affiliation{Fundaci\'on ARAID, Diputaci\'on General de Arag\'on, Zaragoza, Spain}
\author{A.~Maiorano}\affiliation{Dipartimento di Fisica, Sapienza Universit\`a di Roma, Istituto Nazionale di Fisica Nucleare, Sezione di Roma I, I-00185 Rome, Italy}\affiliation{Instituto de Biocomputaci\'on y F\'{\i}sica de Sistemas Complejos (BIFI), 50009 Zaragoza, Spain}
\author{E.~Marinari}\affiliation{Dipartimento di Fisica, Sapienza Universit\`a di Roma, Istituto Nazionale di Fisica Nucleare, Sezione di Roma I, I-00185 Rome, Italy}
\affiliation{Nanotec-Consiglio Nazionale delle Ricerche, I-00185 Rome, Italy}
\author{V.~Martin-Mayor}\affiliation{Departamento  de F\'\i{}sica Te\'orica I, Universidad Complutense, 28040 Madrid, Spain}\affiliation{Instituto de Biocomputaci\'on y F\'{\i}sica de Sistemas Complejos (BIFI), 50009 Zaragoza, Spain}
\author{J.~Monforte-Garcia}\affiliation{Instituto de Biocomputaci\'on y F\'{\i}sica de Sistemas Complejos (BIFI), 50009 Zaragoza, Spain}
\author{A.~Mu\~noz-Sudupe}\affiliation{Departamento  de F\'\i{}sica Te\'orica I, Universidad Complutense, 28040 Madrid, Spain}\affiliation{Instituto de Biocomputaci\'on y F\'{\i}sica de Sistemas Complejos (BIFI), 50009 Zaragoza, Spain}
\author{D.~Navarro}\affiliation{Departamento de Ingenier\'{\i}a, Electr\'onica y Comunicaciones and I3A, U. de Zaragoza, 50018 Zaragoza, Spain}
\author{G.~Parisi}\affiliation{Dipartimento di Fisica, Sapienza Universit\`a di Roma, Istituto Nazionale di Fisica Nucleare, Sezione di Roma I, I-00185 Rome, Italy}
\affiliation{Nanotec-Consiglio Nazionale delle Ricerche, I-00185 Rome, Italy}
\author{S.~Perez-Gaviro}\affiliation{Centro Universitario de la Defensa, Carretera de Huesca s/n, 50090 Zaragoza, Spain}\affiliation{Instituto de Biocomputaci\'on y F\'{\i}sica de Sistemas Complejos (BIFI), 50009 Zaragoza, Spain}
\author{F.~Ricci-Tersenghi}\affiliation{Dipartimento di Fisica, Sapienza Universit\`a di Roma, Istituto Nazionale di Fisica Nucleare, Sezione di Roma I, I-00185 Rome, Italy}
\affiliation{Nanotec-Consiglio Nazionale delle Ricerche, I-00185 Rome, Italy}
\author{J.J.~Ruiz-Lorenzo}\affiliation{Departamento de F\'{\i}sica and Instituto de Computaci\'on Cient\'{\i}fica Avanzada (ICCAEx), Universidad de Extremadura, 06071 Badajoz, Spain}\affiliation{Instituto de Biocomputaci\'on y F\'{\i}sica de Sistemas Complejos (BIFI), 50009 Zaragoza, Spain}
\author{S.F.~Schifano}\affiliation{Dipartimento di Matematica e Informatica, Universit\`a di Ferrara e INFN, Sezione di Ferrara, Ferrara, Italy}
\author{B.~Seoane}\affiliation{Laboratoire de Physique Th\'eorique, \'Ecole Normale Sup\'erieure \& Universit\'e de Recherche Paris Sciences et Lettres, Pierre et Marie Curie \& Sorbonne Universit\'es, UMR 8549 CNRS, 75005 Paris, France}\affiliation{Instituto de Biocomputaci\'on y F\'{\i}sica de Sistemas Complejos (BIFI), 50009 Zaragoza, Spain}
\author{A.~Tarancon}\affiliation{Departamento  de F\'\i{}sica Te\'orica, Universidad de Zaragoza, 50009 Zaragoza, Spain}\affiliation{Instituto de Biocomputaci\'on y F\'{\i}sica de Sistemas Complejos (BIFI), 50009 Zaragoza, Spain}
\author{R.~Tripiccione}\affiliation{Dipartimento di Fisica e Scienze della Terra, Universit\`a di Ferrara e INFN, Sezione di Ferrara, Ferrara, Italy}
\author{D.~Yllanes}\affiliation{Department of Physics and Soft Matter Program, Syracuse University, Syracuse, NY, 13244}\affiliation{Instituto de Biocomputaci\'on y F\'{\i}sica de Sistemas Complejos (BIFI), 50009 Zaragoza, Spain}
\collaboration{Janus Collaboration}
\date{\today}

\begin{abstract}
The unifying feature of glass formers (such as
  polymers, supercooled liquids, colloids, granulars, spin glasses,
  superconductors, ...) is a sluggish dynamics at low
  temperatures. Indeed, their dynamics is so slow that thermal
  equilibrium is never reached in macroscopic samples: in analogy with
  living beings, glasses are said to age.  Here, we show how to relate
  experimentally relevant quantities with the experimentally
  unreachable low-temperature equilibrium phase.
 We have performed a very accurate computation
of the non-equilibrium fluctuation-dissipation ratio for
the three-dimensional Edwards-Anderson Ising spin glass, by means
of large-scale simulations on the special-purpose computers
Janus and Janus II. This ratio (computed for finite times
on very large, effectively infinite, systems) is
compared with the equilibrium probability distribution
of the spin overlap for finite sizes. The resulting quantitative statics-dynamics
dictionary, based on observables that can be measured with current experimental methods,
could allow the experimental exploration of important features
of the spin-glass phase  without  uncontrollable
extrapolations to infinite times or system sizes. 
\end{abstract}
\maketitle

Theory and Experiment follow apparently diverging paths when
studying the glass transition. On the one hand, experimental glass formers
(spin glasses, fragile molecular glasses, polymers, colloids, \ldots) undergo
a dramatic increase of characteristic times when cooled down to their glass
temperature, $T_\mathrm{g}$~\cite{cavagna:09}. Below $T_\mathrm{g}$, the glass
is always out of equilibrium and {\em aging}
appears~\cite{vincent:97}. Consider a rapid quench from a high temperature to
the working temperature $T$ ($T<T_\mathrm{g}$), where the system is left to
equilibrate for time $t_\mathrm{w}$ and probed at a later time
$t+t_\mathrm{w}$.  Response functions such as the magnetic susceptibility turn
out to depend on $t/t_\mathrm{w}^\mu$, with $\mu\approx
1$~\cite{vincent:97,rodriguez:03,dupuis:05}.  The age of the glass,
$t_\mathrm{w}$, remains the relevant time scale even for $t_\mathrm{w}$ as
large as several days.  Relating the aging experimental responses to
equilibrium properties is an open problem.

A promising way to fill the gap is to establish a statics-dynamics dictionary
(SDD)~\cite{barrat:01,janus:08b,janus:10,janus:10b}: non-equilibrium
properties at \emph{finite times} $t$, $t_\mathrm{w}$, as obtained on samples
of macroscopic size $L\to\infty$, are quantitatively matched to equilibrium
quantities computed on systems of \emph{finite size} $L$ [the SDD is an
$L\leftrightarrow (t,t_\mathrm{w})$ correspondence]. Clearly, in order for it to be of
any value, an SDD cannot strongly depend on the particular pair of aging and
equilibrium quantities that are matched.

Some time ago, we proposed one such a
SDD~\cite{janus:08b,janus:10,janus:10b}. However, this SDD was
unsatisfactory in two respects. First, $L$ was matched only to
$t_\mathrm{w}$ (irrespectively of the probing time
$t+t_\mathrm{w}$). Second, our SDD matched spatial correlation
functions whose experimental study is only
incipient~\cite{oukris:10,komatsu:11}.

One could think~\cite{barrat:01} of building an SDD through the Generalized
Fluctuation Dissipation relations (GFDR) first introduced in
\cite{cugliandolo:93} (for related developments see
\cite{franz:95,marinari:98f,franz:98,franz:99,marinari:00b,herisson:02,cruz:03,herisson:04}).
The GFDR are correct at very large times. However, on time scales that can be
investigated in experiments, glassy systems are not fully thermalized since
the approach to equilibrium is very slow. Strong corrections pollute GFDR at
finite times. Here we show how the SDD can be used in a particular case to
compute such corrections (that will be likely present in all glassy systems).
We find that the naive implementation of this idea~\cite{barrat:01} does not
work in general, and we introduce a modified SDD that works for spin glasses
(and, hopefully, also for glasses).

GFDR carry crucial
information~\cite{cugliandolo:93,franz:98,franz:99}: they provide a
promising experimental path towards measuring Parisi's functional
order parameter~\cite{parisi:79}. As a consequence GFDR
have attracted much attention.  One encounters numerical
studies for both Ising~\cite{marinari:98f,marinari:00b,cruz:03} and
Heisenberg~\cite{kawamura:03,billoni:05} spin glasses, as well as for
structural
glasses~\cite{parisi:97b,barrat:99,barrat:00,berthier:07,gnan:13}. On
the experimental side, we have studies on atomic
spin glasses~\cite{herisson:02,herisson:04}, superspin
glasses~\cite{komatsu:11}, polymers~\cite{grigera:99,oukris:10},
colloids~\cite{bellon:01,maggi:10,maggi:12,gomez-solano:09,jop:09,greinert:06,bonn:03}
or DNA~\cite{dieterich:15}.

Here, we perform a detailed simulation of GFDR in the
three-dimensional Ising spin glass employing the custom-made supercomputers
Janus \cite{janus:08} and Janus II \cite{janus:14}. In fact, this has been the
launching simulation campaign of the Janus~II machine, which was designed with
this sort of dynamical studies in mind. Our simulations stand out by the
spanned time range (11 orders of magnitude), by our high statistical accuracy
and by the range of system sizes, enabling us to control size effects
($L=20,\ 40,\ 80$ and $160$). Thus armed, we assess whether or not an SDD can
be built from the GFDR, and compare the SDD proposed in this paper
with other proposals. We focus on spin glasses, rather than on other model
glasses, for a number of reasons: (i) their sluggish dynamics is known to be
due to a thermodynamic phase transition at
$T_\mathrm{c}=T_\mathrm{g}$~\cite{gunnarsson:91,palassini:99,ballesteros:00};
(ii) the linear size of the magnetically correlated domains, $\xi(t_\mathrm{w})$, is
experimentally accessible~\cite{joh:99,bert:04} ($\xi\sim 100$ lattice
spacings~\cite{joh:99}, much larger than comparable measurements for
structural glasses~\cite{berthier:05}); (iii) a GFDR-based SDD has been well
established in the limit of large sizes and
times~\cite{cugliandolo:93,franz:98,franz:99}, see~\eqref{eq:FDTV-1} below;
(iv) GFDR have been studied experimentally~\cite{herisson:02}; (v)
well developed, yet mutually contrasting, theoretical scenarios are available
for spin glasses in equilibrium~\cite{young:98}; (vi) magnetic systems are
notably easier to model and to simulate numerically (in fact, special-purpose
computers have been built for the simulation of spin glasses
\cite{cruz:01,ogielski:85,janus:08, janus:09,janus:14}).


\paragraph*{GFDR and the SDD}
We suddenly cool a three-dimensional spin-glass sample of size $L^3$
from high temperature to the working (sub-critical) temperature
$T=0.7=0.64 T_\mathrm{c}$ at the initial time $t_\mathrm{w}=0$ (see
Methods, below, for more details and definitions).  During the
non-equilibrium relaxation a coherence length $\xi(t_\mathrm{w}$)
grows~\cite{joh:99,janus:08b,janus:09b}, which is representative of
the size of the spin-glass domains. Then, from the waiting time
$t_\mathrm{w}$ on, we place the system under a magnetic field of
strength $H$, and consider the response function at a later measuring
time $t+t_\mathrm{w}$
\begin{equation}\label{eq:lin_susc}
\chi_L(t+\tw,t_\mathrm{w})=\left.\frac{\partial m_L(t+t_\mathrm{w})}{\partial
  H}\right|_{H=0},
\end{equation}
where $m_L(t+t_\mathrm{w})$ is the magnetization density in a sample of linear size $L$.
This susceptibility is then compared with the spin temporal
correlation function $C_L(t+\tw,\tw)$. 
From now on, we shall take
the limits
\begin{align}\label{eq:limite-1}
\chi(t+\tw,t_\mathrm{w})&=\lim_{L\to\infty} \chi_L(t+\tw,t_\mathrm{w})\,,\\
C(t+\tw,t_\mathrm{w})&=\lim_{L\to\infty} C_L(t+\tw,t_\mathrm{w})\,,
\end{align}
which are easy to control numerically: if $L\gtrsim 7 \xi(t+t_\mathrm{w})$
size effects are negligible~\cite{janus:08b}\footnote{In fact, the correlation
  functions decay exponentially with distance. Therefore, with periodic
  boundary conditions, size effects should decay exponentially with
  $L/\xi$. Indeed, an explicit computation shows that, to our accuracy level,
  size corrections are completely negligible when $L>7\xi$~\cite{janus:08b}.}
(see also Appendix~\ref{sec:finite-size}).

The Fluctuation Dissipation Theorem (FDT)
states that $T \chi(t+\tw,t_\mathrm{w})=1-C(t+\tw,t_\mathrm{w})$, with both
$\chi$ and $C$ computed at $H=0$. However, for $T<T_\mathrm{c}$ the FDT does
not hold. In fact, GFDR
take the form~\cite{cugliandolo:93,franz:98,franz:99} (the
order of limits is crucial):
\begin{equation}
\lim_{t_\mathrm{w}\to\infty} T\chi(t+\tw,t_\mathrm{w})= \lim_{t_\mathrm{w}\to\infty} [\lim_{L\to\infty} S(C_L(t+\tw,t_\mathrm{w}),L)]\,,\label{eq:FDTV-1}
\end{equation}
where $t$ is scaled as $t_\mathrm{w}$ grows, to ensure that the full range
$0<C(t+\tw,t_\mathrm{w})<1$ gets covered, and $S(C,L)$ is given by a double integral of $P(q,L)$, the
equilibrium distribution function of the spin overlap, whose explicit
definition is provided in the Methods Section.

Here, we mimic an experimental protocol~\cite{herisson:02,herisson:04} in that we
consider the non-equilibrium response on a very large system but at
\emph{finite times}. We try to relate this response with the equilibrium
overlap for a system of finite effective size $L_{\mathrm{eff}}$
\begin{equation}\label{eq:FDTV-2}
T\chi(t+\tw,t_\mathrm{w}) =
S\big(C(t+\tw,t_\mathrm{w}),L_{\mathrm{eff}}(t+t_\mathrm{w}, t_\mathrm{w})\big)\,,
\end{equation}
where we have assumed that both $\chi$ and $C$ have reached their
thermodynamic limit. The same approach was followed for a two-dimensional spin
glass by Barrat and Berthier~\cite{barrat:01} (note, however, that there is no
stable spin-glass phase at $T>0$ in two spatial dimensions).

Eq.~\eqref{eq:FDTV-2} provides  a statics-dynamics dictionary (SDD)  relating both
times  $t$  and  $t_\mathrm{w}$  with  a  single  effective  equilibrium  size
$L_{\mathrm{eff}}(t+t_\mathrm{w}, t_\mathrm{w})$. Note that  it is not obvious
a priori that our  program can be carried out. For instance,  our SDD does not
exist for ferromagnets, as explained in details in Appendix~\ref{sec:ferromagnet} exploiting data
from Refs.~\cite{parisi:99b,ricci-tersenghi:03}.

SDDs based on the comparison of aging and equilibrium correlation functions
(rather than on GFDR) have been studied in some
detail~\cite{janus:10,janus:10b,fernandez:15}. It was found that
the effective length depends solely on $t_\mathrm{w}$. Indeed,
\begin{equation}\label{eq:SDD-simple}
L_{\mathrm{eff}}(t+t_\mathrm{w}, t_\mathrm{w}) = k\, \xi(t_\mathrm{w})\,,
\end{equation}
with $k\approx 3.7$, was accurate enough to match the correlation
functions~\cite{janus:10,janus:10b}. Ref.~\cite{barrat:01} also agreed with
\eqref{eq:SDD-simple}. In fact, \eqref{eq:SDD-simple} also underlies the analysis of Refs. ~\cite{manssen:15b,wittmann:16}. Yet, we shall show below that~\eqref{eq:SDD-simple} is oversimplified.

\begin{figure}[t]
\centering \includegraphics[width=\columnwidth, angle=0]{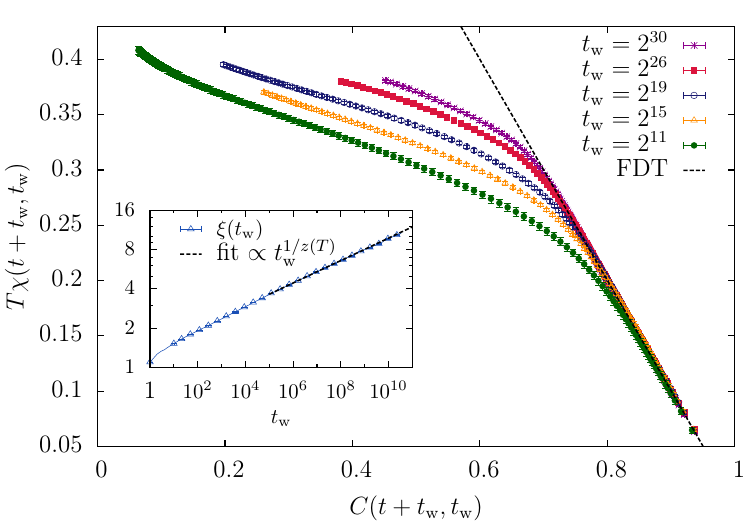}
\caption{Response function $T \chi(t+\tw,t_\mathrm{w})$
  versus $C(t+\tw,t_\mathrm{w})$ at $T=0.7$ [for fixed $t_\mathrm{w}$,
  $C(t+\tw,t_\mathrm{w})$ monotonically decreases from $C=1$ at $t=0$ to
  $C=0$ at $t=\infty$]. Data for $t_\mathrm{w}=2^{11}$ and
  $t_\mathrm{w}=2^{30}$ were obtained on Janus II (the other $\tw$ are from
  Janus). The five values of $t_\mathrm{w}$ correspond to effective
  equilibrium sizes $L_\mathrm{eff}$ that, according to
    \eqref{eq:SDD-simple}, span the size range investigated in
  Ref.~\cite{janus:10} (namely, $8\leq L \leq 32$). {\bf Inset:}
  growth of the spin-glass coherence length $\xi(t_\mathrm{w})$ as a
  function of time, computed at zero magnetic field and following
  Refs.~\cite{janus:08b,janus:09b}, from simulations of $L=160$
  lattices at $T=0.7$ on Janus II. In dashed lines we plot the scaling
  $\xi(t_\mathrm{w})\propto t_\mathrm{w}^{1/z(T)}$ with $z(T)=11.64$
  from Ref.~\cite{janus:09}.}
\label{fig:FDT-xi}
\end{figure}

\paragraph*{Numerical data} 
The three basic quantities computed in this work, namely
$\chi(t+\tw,t_\mathrm{w}), C(t+\tw,t_\mathrm{w})$ and $\xi(t_\mathrm{w})$
are displayed in  Fig.~\ref{fig:FDT-xi}. Full details about this computation
are provided in Appendix~\ref{sec:susclin}.

Let us remark that the Janus II supercomputer allows us to probe
unexplored dynamical regimes, either $t/t_\mathrm{w}$ as large as
$2^{24}\approx 1.4
\times 10^{7}$ (i.e., we follow the magnetic response for a very long
time, after the field was switched on at $t_\mathrm{w}=2^{11}$) or
$t_\text{w}$ as large as $2^{30}$ (i.e., we study the response of a
very old spin glass, but we are limited to $t/t_\mathrm{w}\approx 27$
in this case).

It is also remarkable that we are able to compute both the
susceptibility $\chi$ and the correlation function $C$ without
worrying about finite-size effects. Indeed, size effects become
visible when the coherence length reaches the threshold
$\xi(t_\mathrm{w})\approx L/7$~\cite{janus:08b} which in our $L=160$ lattice
translates to $\xi\approx 23$ lattice spacings. As
Fig.~\ref{fig:FDT-xi}--inset shows, we are quite far from this safety
threshold.

With respect to previous measurements of the GFDR ratio, it is worth
stressing that now we are able to take the $h\to 0$ limit in a more
controlled way. This is far from trivial, given that the linear
response regime shrinks to very small field when $\tw$ increases (see
Appendix~\ref{sec:susclin}).

The data in Fig.~\ref{fig:FDT-xi} also stand out by their statistical
accuracy (due to the large number of samples and large system sizes we
simulated, but also thanks to the analysis method described in Appendix~\ref{sec:smooth}
As a consequence, a behavior different from the one implied by
FDT,
$T \chi(t,t_\mathrm{w})=1-C(t,t_\mathrm{w})$ can be studied in
detail. In particular, the reader might be stricken by the linear
behavior at $C(t+\tw,\tw)\approx 0.4$. In fact, following
Refs.~\cite{cugliandolo:93,franz:98,franz:99}, this linear behavior
could be interpreted as evidence for one step of replica-symmetry
breaking (see, for instance, Ref.~\cite{mezard:87}). However, we shall
argue below that the effective length in \eqref{eq:FDTV-2} evolves as
time $t$ grows, thus producing an upturn in the response which is probably
responsible for the linear behavior in Fig.~\ref{fig:FDT-xi}.

Let us make a final remark. We know that $S(C,L)$ is upper bounded by $1 -
\overline{\langle |q| \rangle }_{L=\infty} \geq 1 -q_\mathrm{EA}^{(L=\infty)}$
(see Methods for definitions; the proof of the inequality is outlined in Appendix~\ref{sec:inequality}. At $T=0.7$ we know that $1
-q_\mathrm{EA}^{(L=\infty)}=0.48(3)$~\cite{janus:10b} (or
0.46(3)~\cite{janus:10}).  Therefore, the dynamic responses
$T\chi(t,t_\mathrm{w})$ in Fig.~\ref{fig:FDT-xi} are well below $1
-q_\mathrm{EA}^{(L=\infty)}$ and \eqref{eq:FDTV-2} could be satisfied. The
general conditions under which \eqref{eq:FDTV-2} can be used are discussed in the Appendix.

\begin{figure}[t]
\centering \includegraphics[width=\columnwidth, angle=0]{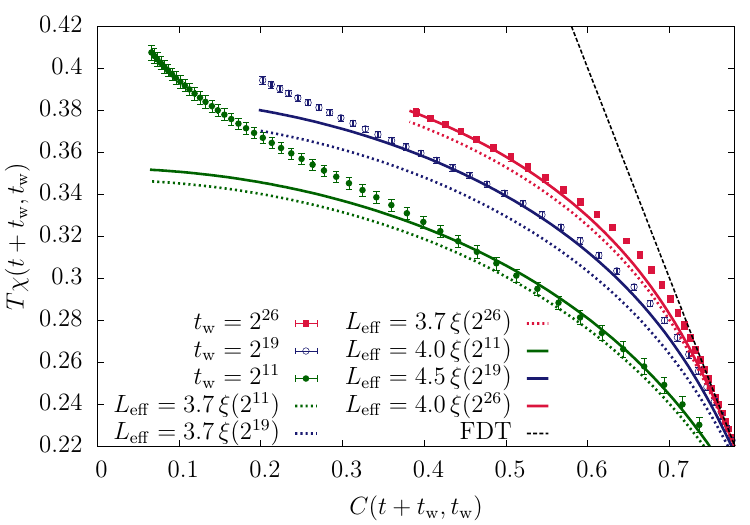}
\caption{Close-up of Fig.~\ref{fig:FDT-xi} (we only
  show data for three $t_\mathrm{w}$, for the sake of clarity). Lines are
  $S(C,L_{\mathrm{eff}})$, recall \eqref{eq:FDTV-2}, with the
  effective equilibrium size as in \eqref{eq:SDD-simple}:
  $L_{\mathrm{eff}}(t+t_\mathrm{w}, t_\mathrm{w}) =
  k\xi(t_\mathrm{w})$. Dotted lines correspond to $k=3.7$, which is
  the proportionality constant that was found by matching equilibrium
  and non-equilibrium correlation
  functions~\cite{janus:08b,janus:10,janus:10b}. The continuous lines
  were found by choosing the best possible $k$
  for each $t_\mathrm{w}$. This
representation shows that the single-time statics-dynamics dictionary
$L_\text{eff}\sim \xi(\tw)$ breaks down for large $t$, when $\xi(t+\tw)$
is much larger than $\xi(\tw)$.
\label{fig:FDT-no-one-time}}
\end{figure}

\begin{figure}[t]
\centering \includegraphics[width=\columnwidth, angle=0]{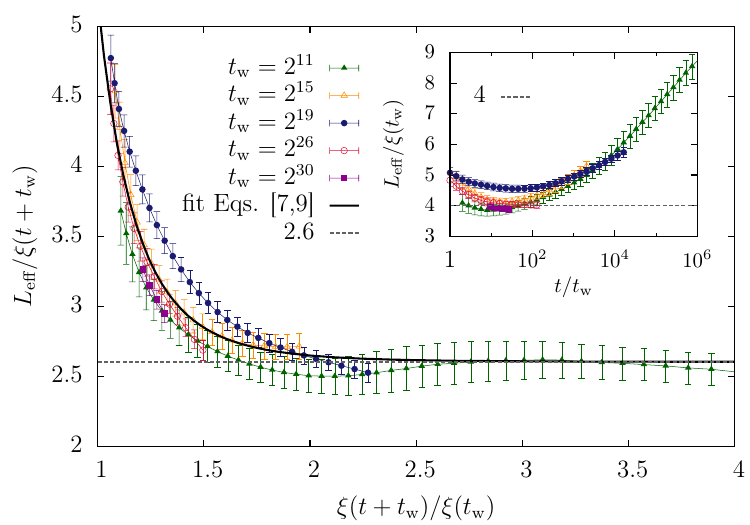}
\caption{For each $t_\mathrm{w}$, we show the effective
  equilibrium size $L_{\mathrm{eff}}(t+t_\mathrm{w}, t_\mathrm{w})$ in
  units of the coherence length at the measuring time
  $\xi(t+t_\mathrm{w})$ versus the ratio of coherence lengths
  $\xi(t+t_\mathrm{w})/\xi(t_\mathrm{w})$ (recall that $t$ is the time
  elapsed since switching-on the magnetic field). The ratio of
  coherence lengths is 1 for $t=0$ and goes as
  $\xi(t+t_\mathrm{w})/\xi(t_\mathrm{w})\propto
  (1+t/t_\mathrm{w})^{1/z(T)}$ for large time, with
  $z(T=0.7)=11.64(15)$~\cite{janus:09b}. Let us stress that there is
  no extrapolation in this figure, only interpolation (i.e.,
  $L_\text{eff}$ falls within the simulated equilibrium sizes, $8\leq
  L_\mathrm{eff}\leq 32$). The solid line is a fit to the scaling
  function $h(x)$ in \eqref{eq:ultrametric-ansatz} and
  \eqref{eq:simpler-ansatz}. {\bf Inset:}
  $L_{\mathrm{eff}}(t+t_\mathrm{w}, t_\mathrm{w})$ data from the main
  panel in units of the coherence length at the initial time time
  $\xi(t_\mathrm{w})$, as a function of the time ratio
  $t/t_\mathrm{w}$.}
\label{fig:FDT-Leff}
\end{figure}

\paragraph*{The effective equilibrium size} As we show in
Fig.~\ref{fig:FDT-no-one-time}, our data are too accurate to be
quantitatively described by combining \eqref{eq:FDTV-2} with
\eqref{eq:SDD-simple}. This simple description fails both at short
times $t$ (i.e., when $C(t,t_\mathrm{w})\approx q_\mathrm{EA}^{[L\approx
  4\xi(t_\mathrm{w})]}$) and also at very long $t$, although one can
find a constant $k$ that works well for intermediate $t$.

The discrepancy for long $t$ seems easy to rationalize: since the
growth of $\xi(t_\mathrm{w})$ is very slow, recall
Fig.~\ref{fig:FDT-xi}--inset, $\xi(t+t_\mathrm{w})$ and
$\xi(t_\mathrm{w})$ are very similar to each other for small $t$ and, therefore,
$L_\mathrm{eff}\propto \xi(t_\mathrm{w})$ makes sense. However, since
$\xi(t_\mathrm{w})$ grows without bounds in the spin-glass phase, one should
eventually have $\xi(t+t_\mathrm{w}) \gg \xi(t_\mathrm{w})$. Under these
circumstances, it is only natural that
$L_\mathrm{eff}\propto\xi(t+t_\mathrm{w})$. 

We can test this proposal by computing an exact $L_\text{eff}$
for each $(t,t_\mathrm{w})$ pair (see Appendix~\ref{sec:Leff} for
details), which we plot in Fig.~\ref{fig:FDT-Leff}: in the main panel
in units of $\xi(t+\tw)$ and in the inset in units of $\xi(\tw)$.

The first important observation from the main panel in
Fig.~\ref{fig:FDT-Leff} is that, for long enough times, we find
$L_{\mathrm{eff}}\approx 2.6\,\xi(t+t_\mathrm{w})$, in agreement with
the intuition exposed above.  This is definitely different from
\eqref{eq:SDD-simple}, used until now.  The data in the inset of
Fig.~\ref{fig:FDT-Leff} explain why the previous relation in
\eqref{eq:SDD-simple} passed many numerical tests until now: the
non-monotonic behavior of $L_\mathrm{eff}/\xi(\tw)$ for short times
$t$ makes this ratio roughly compatible with a constant $k \approx 4$
as long as $t/\tw \lesssim 1000$.

Surprisingly, the ratio $L_\mathrm{eff}/\xi(t+\tw)$, or equivalently
$L_\mathrm{eff}/\xi(\tw)$, becomes large as well when $t\to 0$,
thus explaining the inability of \eqref{eq:FDTV-2} in describing
dynamical data at short times $t$ (see
Fig.~\ref{fig:FDT-no-one-time}).  Nonetheless in the limit $t\to 0$,
i.e.\ $\xi(t+\tw)/\xi(\tw)\to 1$, the effective equilibrium size
$L_\mathrm{eff}$ seems to reach a finite value; a divergence of
$\Leff$ in this limit seems unlikely (see Appendix~\ref{sec:Pdyn}).

\paragraph*{$L_\mathrm{eff}$ and the spin-glass coherence length} 
Now that it is clear that both $\xi(t_\mathrm{w})$ and
$\xi(t+t_\mathrm{w})$ are relevant for $L_\mathrm{eff}$ one may ask
about the crossover between the $\xi(t_\mathrm{w})$-dominated regime
and the $\xi(t+t_\mathrm{w})$-dominated regime. Fig.~\ref{fig:FDT-Leff} tells
us that $L_\mathrm{eff}/\xi(t+t_\mathrm{w})$ is, to a good approximation, a
function of the ratio $\xi(t+t_\mathrm{w})/\xi(t_\mathrm{w})$.\footnote{The
reader will note that data for $t_\mathrm{w}=2^{19}$ are slightly off, in
Fig.~\ref{fig:FDT-Leff}. We attribute the effect to a strong statistical
fluctuation, enhanced by the fact that all data points with the same
$t_\mathrm{w}$ are extremely correlated.} Thus, we attempted to fit the
crossover with the functional form
\begin{equation}\label{eq:ultrametric-ansatz}
L_\mathrm{eff}(t+t_\mathrm{w},t_\mathrm{w})=\xi(t+t_\mathrm{w})\,
h\big(\xi(t+t_\mathrm{w})/\xi(t_\mathrm{w})\big)\,,
\end{equation}
where the scaling function is
\begin{equation}\label{eq:almost-ultrametric-ansatz}
h(x)=k_1\ +\ k_2\, x^{-c}\,.
\end{equation}
Interpolation of data shown in Fig.~\ref{fig:FDT-Leff} returns:
$k_1=2.58(2)$, $k_2=2.7(1)$ and $c=5.9(2)$.
Noticing that $k_2 \approx k_1$ and $c \approx z(T)/2$, where $z(T)$ is the exponent for the time
growth of the coherence length, $z(T=0.7)=11.64(15)$ (see Fig.~\ref{fig:FDT-xi}--inset, and
Refs.~\cite{janus:08b,janus:09b}), the scaling function $h(x)$
can be also rewritten in a much simpler form as
\begin{equation}
h\big(\xi(t+t_\mathrm{w})/\xi(t_\mathrm{w})\big) = k_1 \left(1 + \sqrt{\frac{\tw}{t+\tw}}\right)
\label{eq:simpler-ansatz}
\end{equation}
Fitting data in Fig.~\ref{fig:FDT-Leff} with this simpler scaling function
returns $k_1 = 2.59(1)$ (see full curve in Fig.~\ref{fig:FDT-Leff}).
Given that the fit with 3 adjustable parameters in \eqref{eq:almost-ultrametric-ansatz}
and the one in \eqref{eq:simpler-ansatz} with just 1 adjustable parameter
have practically the same quality-of-fit, we tend to prefer the simpler
ansatz, as long as it interpolates the numerical data well enough.

\begin{figure}[t]
\centering \includegraphics[width=\columnwidth, angle=0]{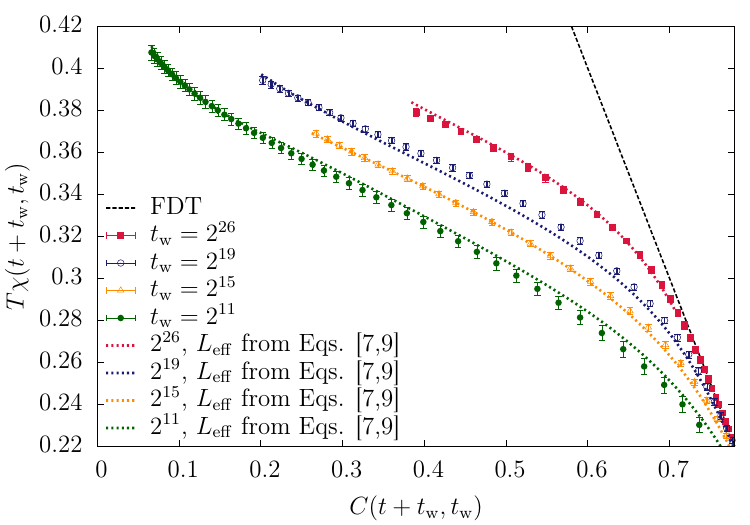}
\caption{As in Fig.~\ref{fig:FDT-no-one-time} but
  $L_\mathrm{eff}$ is taken from the ansatz in
  \eqref{eq:ultrametric-ansatz} and \eqref{eq:simpler-ansatz}, which
  improves on the single-time statics-dynamics dictionary based on
  $\xi(\tw)$ by considering a crossover to a $\xi(t+\tw)$-dominated
  regime.
\label{fig:dominance-large-power}}
\end{figure}

The ultimate check for the success of
\eqref{eq:ultrametric-ansatz} and \eqref{eq:simpler-ansatz} in
reproducing the aging response is provided by
Fig.~\ref{fig:dominance-large-power}, where the dynamical measurements
(data points with errors) are plotted together with the equilibrium
function $S(C(t+\tw,\tw),\Leff(t+\tw,\tw))$.  The very good agreement
in the whole range gives a strong support in favor of an SDD based on
\eqref{eq:ultrametric-ansatz} and \eqref{eq:simpler-ansatz}.

Note as well that \eqref{eq:ultrametric-ansatz} explains the
previous success of the simpler SDD in \eqref{eq:SDD-simple}. In
fact, at short times $t$, the two coherence lengths
$\xi(t+t_\mathrm{w})$ and $\xi(t_\mathrm{w})$ are very similar to each
other, and the amplitude $k$ in \eqref{eq:SDD-simple} is
essentially $k=k_1+k_2\approx 2 k_1$.

The ansatz of~\eqref{eq:ultrametric-ansatz} provides as well a simple
explanation for the upturn of the aging response at small values of
$C$, recall Fig.~\ref{fig:FDT-xi}.  Indeed, as time $t$
increases, the correlation function decays as
$C\propto(t+t_\mathrm{w})^{-1/\alpha}, \alpha\approx
7$~\cite{janus:08b}.  But, from $\xi(t+t_\mathrm{w})\propto
(t+t_\mathrm{w})^{1/z(T)}$ we conclude that, even at fixed
$t_\mathrm{w}$, $L_\mathrm{eff}$ diverges for large $t$ as
$C^{-\alpha/z(T)}$. Now, to a first approximation, one may expect that
$S(C,L=\infty)-S(C,L)\propto L^{-\theta\approx -0.38}$ (see the
description of the overlap distribution function in the Methods
section, below). We thus expect the susceptibility to approach its
$C=0$ limit in a singular way, as $C^{\theta/(\alpha z(T))}\approx C^{0.23}$.

\paragraph*{Which features of the $P(q)$ can be obtained from dynamic measurements?} 
One of the major gains of the present analysis would be to obtain
Parisi's functional order parameter $P(q)$ from experimental dynamic
data. In an ideal situation, one would have data for
$\chi$, $C$ and $\xi$, complemented by the ansatz
in~\eqref{eq:simpler-ansatz}. Then, one would like to know which
features of the underlying $S(C,L)$ can be retrieved from these
dynamic measurements.

In order to answer this question, we have considered a very simplified
$P_\mathrm{simpl}(q,L)$, that possesses the main features of the
$P(q,L)$ measured in numerical simulations (see Methods):
\begin{align}
P_\mathrm{simpl}(q,L) = &
\,\big(P_0 + P_1 q^2\big)\,\mathbbm{1}\big[|q| < q_\mathrm{EA}^{(L)}\big]+\nonumber\\
&w^{(L)}\big(\delta(q-q_\mathrm{EA}^{(L)})+\delta(q+q_\mathrm{EA}^{(L)})\big)/2\,,\label{eq:Pq-syn}
\end{align}
where $P_0$ and $P_1$ are constants, $\mathbbm{1}$ is the indicator function
and $w^{(L)}$ is a weight enforcing normalization.\footnote{Note that the delta peak in~\eqref{eq:Pq-syn} is a reasonable expectation only for an
  infinite system (see Methods).}  Integrating 
$P_\mathrm{simpl}(q,L)$ twice we get
\begin{equation}\label{eq:S-syn}
S_\mathrm{simpl}(C,L)=\min\left[S_0(L)- P_0 C^2-\frac{P_1}{6} C^4, 1-C\right]\,.
\end{equation}
We take $S_0(L)=S(0,L)$ from the true $P(q,L)$. Recall that
$S(0,L)=1-\overline{\langle |q|\rangle}_L$, see Appendix~\ref{sec:inequality}.  Instead, the
$L$-independent $P_0$ and $P_1$ are fitted in order to obtain a
$S_\mathrm{simpl}(C,L)$ as similar as possible to the true
$S(C,L)$: we get $P_0=0.167(1)$ and $P_1=0.46(3)$.
In other words, $P_\mathrm{simpl}(q)$ shares with the true
distribution only four numeric features: normalization, first absolute
moment $\overline{\langle |q|\rangle}_L$, $P_0 \simeq P(q=0,L)$ which
is essentially $L$-independent, and the second derivative $P_1 \simeq
P''(q=0,L)/2$. In particular, note that having $P_0>0$ is a crucial
feature of the mean-field solution~\cite{marinari:00}.
A direct measure for sizes $8\le L \le 32$ returns the
$L$-independent value $P(q=0,L) = 0.167(5)$ \cite{janus:10} confirming the
validity of our simplified description.

\begin{figure}[t]
\centering \includegraphics[width=\columnwidth, angle=0]{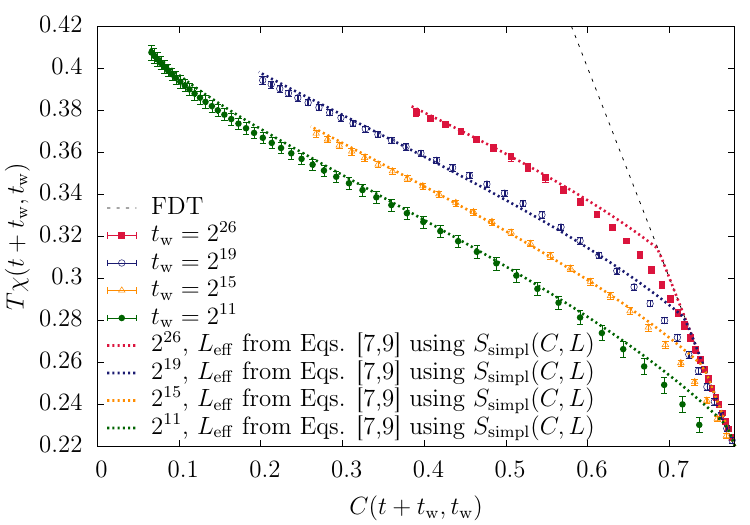}
\caption{As in Fig.~\ref{fig:dominance-large-power} but
  this time we use the simplified $S_\mathrm{simpl}(C,L)$ from
  \eqref{eq:S-syn}. Note that dynamic data are well reproduced
  by~\eqref{eq:ultrametric-ansatz} and \eqref{eq:simpler-ansatz}, even
  in this simple approximation.
\label{fig:dominance-large-power-syn}}
\end{figure}

The outcome of this analysis is given in
Fig.~\ref{fig:dominance-large-power-syn}. It turns out that the
simplified $S_\mathrm{simpl}$ in \eqref{eq:S-syn} is almost as
effective as the true $S(C,L)$ in representing the non-equilibrium
data through the effective size $L_\mathrm{eff}$ in~\eqref{eq:simpler-ansatz}.
The only obvious disagreement is that
\eqref{eq:S-syn} predicts a non-analytic behavior for the
susceptibility $\chi$ at $C=q_\mathrm{EA}^{(L_\mathrm{eff})}$, which is
not found in the non-equilibrium data. In other words, the effective
size for times such that $C(t+t_\mathrm{w},t_\mathrm{w})\approx
q_\mathrm{EA}^{(L\approx 4\xi(t_\mathrm{w}))}$ is large, but certainly
$L_\mathrm{eff}$ is not infinite as demanded by \eqref{eq:Pq-syn}.

Fortunately, even the crude description in~\eqref{eq:S-syn} could lead
to some interesting analysis. For instance, one could select pairs of
times $(t,t_\mathrm{w})$ such that
$L_\mathrm{eff}(t+t_\mathrm{w},t_\mathrm{w})=constant$. Then, $S(0,L_\mathrm{eff})$
will be the same for all those points. Now, we note
from~\eqref{eq:simpler-ansatz} that $\xi(t+t_\mathrm{w})$ can vary by as
much as a factor of two, for such points. It follows that
$C(t+t_\mathrm{w},t_\mathrm{w})$ should vary significantly over this set of times
with fixed $L_\mathrm{eff}(t+t_\mathrm{w},t_\mathrm{w})$. Hence, the crucial
parameters $P_0$ and $P_1$ could be extracted. For instance, if the
susceptibility $\chi(t+t_\mathrm{w},t_\mathrm{w})$ would turn out not to depend on
$C(t+t_\mathrm{w},t_\mathrm{w})$ (while keeping $L_\mathrm{eff}$ fixed), this would
mean $P_0,P_1\approx 0$, in contrast with the mean field prediction $P_0>0$.

\paragraph*{Discussion}
It was discovered some twenty years ago that experimental aging
response functions carry information on Parisi's functional order
parameter~\cite{cugliandolo:93,franz:95,marinari:98f}. We now know
that this connection between non-equilibrium and equilibrium physics
relies on a very general mathematical property, stochastic
stability~\cite{franz:98,franz:99},  shared by many glass
models. However, experimental attempts to
explore this connection encountered a major
problem~\cite{herisson:02,herisson:04}: an essentially uncontrolled
extrapolation to infinite waiting time $t_\mathrm{w}$ is
required.\footnote{See Ref.~\cite{joh:96} for an experimental attempt to
  measure Parisi's functional order parameter, unrelated to GFDR.}

Here, we have proposed employing a statics-dynamics
dictionary~\cite{barrat:01,janus:08b,janus:10,janus:10b} to avoid
uncontrolled extrapolations. Indeed, we have shown that the aging
responses at finite $t_\mathrm{w}$ can be connected to the Parisi's
order parameter as computed at equilibrium in a system of finite size.

We have shown that this GFDR-based SDD is essentially consistent with
previous proposals~\cite{janus:08b,janus:10,janus:10b} that focused on
spatial correlation functions. This is an important consistency
test. There is a caveat, though: when the probing time
$t+t_\mathrm{w}$ is such that one has
$\xi(t+t_\mathrm{w})\gg\xi(t_\mathrm{w})$ for the coherence lengths,
the GFDR-based SDD disagrees from previous dictionaries in that the
size of the equivalent equilibrium system is $L_\mathrm{eff}\sim
\xi(t+t_\mathrm{w})$ (rather than $L_\mathrm{eff}\sim
\xi(t_\mathrm{w})$). In fact, we have found that the $L_\mathrm{eff}$
dependence on both length scales can be simply parameterized, recall
\eqref{eq:ultrametric-ansatz} and \eqref{eq:simpler-ansatz}.

At this point, the reader may wonder about the relationship between
$L_\mathrm{eff}(t+t_\mathrm{w},t_\mathrm{w})$ and the two-time correlation
length $\zeta(t+t_\mathrm{w},t_\mathrm{w})$ obtained from the
two-time/two-site correlation function introduced in
Ref.~\cite{jaubert:07,chamon:07}.  Indeed, we thoroughly studied the
two-time/two-site correlation function in~\cite{janus:09b} because it was a
crucial ingredient for our previous SDD
proposal~\cite{janus:10,janus:10b}. We found (see Fig.~12 in
Ref.~\cite{janus:09b}) that $\zeta(t+t_\mathrm{w},t_\mathrm{w})$ can grow, at
most, as large as $\xi(t_\mathrm{w})$. Instead, the
$L_\mathrm{eff}(t+t_\mathrm{w},t_\mathrm{w})$ introduced here is
asymptotically as large as $\xi(t+t_\mathrm{w})$.

On the other hand, the only previous SDD known to us that was based on
\eqref{eq:FDTV-2} misses the $L_\mathrm{eff}\sim
\xi(t+t_\mathrm{w})$ behavior~\cite{barrat:01}. There are a couple of
possible reasons for this failure. For one, the
time scales in Ref.~\cite{barrat:01} do not allow for length-scale
separation $\xi(t+t_\mathrm{w})\gg\xi(t_\mathrm{w})$. Besides, the SDD
from Ref.~\cite{barrat:01} was obtained for two-dimensional
spin glasses (which only have a paramagnetic phase). Therefore, the
results of Ref.~\cite{barrat:01} are probably a manifestation of
finite-time/finite-size scaling~\cite{lulli:15,fernandez:15}.

Let us conclude by stressing that the three basic quantities analyzed
in this work, namely the susceptibility $\chi(t+\tw,t_\mathrm{w})$,
the correlation function $C(t+\tw,\tw)$ and the coherence length
$\xi(t+t_\mathrm{w})$, have been obtained experimentally in a dynamic
setting very similar to simulations (for $\chi$ and $C$, see
Refs.~\cite{herisson:02,herisson:04}, for $\xi$ see
Refs.~\cite{joh:99,bert:04}). We thus think that it should be possible
to extract the spin-glass functional order parameter from already
existing experimental data. Furthermore, GFDR  have been
studied as well in superspin glasses~\cite{komatsu:11} and in a
variety of soft condensed-matter
systems~\cite{grigera:99,oukris:10,bellon:01,maggi:10,maggi:12,gomez-solano:09,jop:09,greinert:06,bonn:03,dieterich:15}. We
therefore expect that our analysis will be of interest beyond the
realm of spin glasses.
\paragraph*{Acknowledgments --- }
Some of the simulations in this work (the $L<80$ systems, to check for
size effects) where carried out on the \emph{Memento} cluster: we
thank staff from BIFI's supercomputing center for their assistance.
We thank Giancarlo Ruocco for guidance on the experimental literature.
We warmly thank M. Pivanti for his contribution to the early stages of
the development of the Janus II computer. We also thank Link
Engineering (Bologna, Italy) for their precious role in the technical
aspects related to the construction of Janus II. We thank EU,
Government of Spain and Government of Aragon for the financial support
(FEDER) of Janus II development. This work was partially supported by
MINECO (Spain) through Grant Nos. FIS2012-35719-C02, FIS2013-42840-P,
FIS2015-65078-C2, and by the Junta de Extremadura (Spain) through
Grant No. GRU10158 (partially funded by FEDER). This project has
received funding from the European Union's Horizon 2020 research and
innovation program under the Marie Sk{\l}odowska-Curie grant
agreement No. 654971. This project has received funding from the
European Research Council (ERC) under the European Union's Horizon
2020 research and innovation program (grant agreement No 694925).
DY acknowledges support by NSF-DMR-305184 and by the Soft Matter
Program at Syracuse University. MBJ acknowledges the financial support
from ERC grant NPRGGLASS.

\balancecolsandclearpage
\renewcommand{\tocname}{Appendices}
\tableofcontents
\appendix
\section{Model and observables}\label{sec:methods}
We study the $D\!=\!3$ Edwards-Anderson model, whose  Hamiltonian is given by
\begin{equation}
{\cal H}=-\sum_{\langle \boldsymbol{x}, \boldsymbol{y}\rangle } J_{\boldsymbol{x},\boldsymbol{y}} \s_{\boldsymbol{x}}\, \s_{\boldsymbol{y}}-H\sum_{\boldsymbol{x}}\s_{\boldsymbol{x}}\,.\label{eq:EA-H}
\end{equation}
The spins $s_{\boldsymbol{x}}\!=\!\pm1$ are placed on the nodes,
$\boldsymbol{x}$, of a cubic lattice of linear size $L$ and we set
periodic boundary conditions. The couplings
$J_{\boldsymbol{x},\boldsymbol{y}}\!=\!\pm 1$, which join nearest
neighbors only, are chosen randomly with $50\%$ probability and are
quenched variables. For each choice of the couplings (one ``sample''),
we simulate two independent copies of the system,
$\{s_{\boldsymbol{x}}^{(1)}\}$ and $\{s_{\boldsymbol{x}}^{(2)}\}$. We
denote by $\langle\cdots\rangle$ the average over the thermal noise
and by $\overline{(\cdot \cdot \cdot)}$ the \emph{subsequent} average
over the samples.  The model described by \eqref{eq:EA-H} undergoes a
SG transition at $H=0$ and $T_\mathrm{c}=1.102(3)$~\cite{janus:13}.

For our dynamical data we have run new non-equilibrium simulations
on Memento, Janus and Janus II. We use heat-bath dynamics, in which 
one Monte Carlo step roughly corresponds to one picosecond
of the experimental system~\cite{mydosh:93}. See Appendix~\ref{sec:simulations} for 
technical details of these simulations.
The two main dynamical observables are the magnetization density
$m_L(t+t_\mathrm{w})=\overline{\sum_{\boldsymbol{x}}\,\langle
  s_{\boldsymbol{x}}(t+t_\mathrm{w})\rangle}/V$
and the spin temporal correlation function
$C_L(t+\tw,t_\mathrm{w};H)=\overline{\sum_{\boldsymbol{x}}\,\langle
  s_{\boldsymbol{x}}(t_\mathrm{w}) s_{\boldsymbol{x}}(t+t_\mathrm{w})\rangle}/V$.

Equilibrium results at $T=0.7$
are available for $L\leq 8\leq 32$~\cite{janus:10}. 
In this case the main quantity is 
the probability density function $P(q,L)$ of
the spin overlap $q$:
\begin{equation}\label{eq:q}
q\equiv \frac{1}{V} \sum_{\boldsymbol x} s^{(1)}_{\boldsymbol{x}}
s^{(2)}_{\boldsymbol x}\,,\quad
\overline{\langle q^k\rangle}_{L} = \int_{-1}^1 \mathrm{d}q'\  (q')^k P(q',L)\,.
\end{equation}
In particular, we are interested in the integral
\begin{equation}\label{eq:SCL}
S(C,L)=\int_C^1 \mathrm{d}\,C'\, x(C',L)\,,\  x(C,L)=\int_{0}^C\,\mathrm{d} q\, 2 P(q,L)\,.
\end{equation}
The $P(q,L)$ curves are easily described for finite $L$. 
They are symmetric under $q\leftrightarrow
-q$, with two maxima at $\pm q_\mathrm{EA}^{(L)}$ and a flat central region. In
the thermodynamic limit, the two peaks turn into delta functions at $\pm
q_\text{EA}^{(\infty)}$, which mark the maximum possible value of $|q|$.  The size
evolutions, as checked for $L\leq 32$~\cite{janus:10}, are as follows:
$q_\mathrm{EA}^{(L)}-q_\mathrm{EA}^{(\infty)}\propto L^{-\theta\approx0.38}$
(at $T=0.7$, $q_\mathrm{EA}^{(\infty)}=0.52(3)$~\cite{janus:10b}), the width
of the peaks at $\pm q_\mathrm{EA}^{(L)}$ scales as $L^{-B\approx 0.28}$ while
$P(q=0,L)$ turns out to be greater than zero and $L$-independent.

\section{Our  simulations}\label{sec:simulations}
\begin{figure}[t]
 \centering \includegraphics[width=\columnwidth, angle=0]{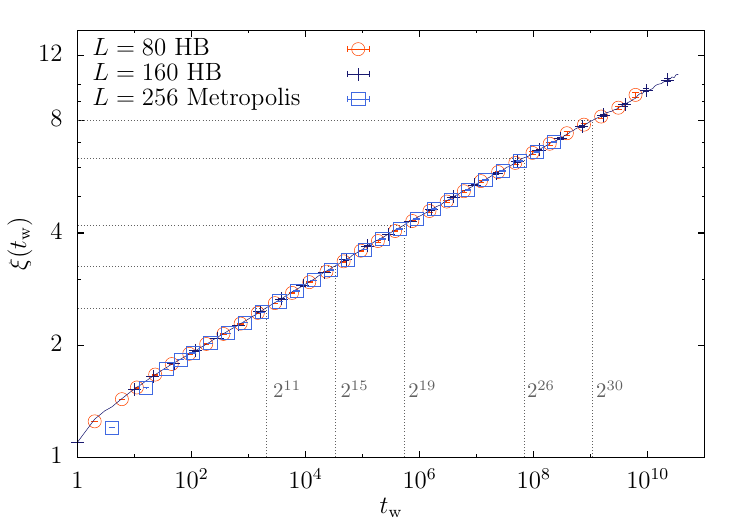}
\caption{Coherence length $\xi(t_\mathrm{w})$ versus
  waiting time $\tw$ at $T=0.7$ for different lattice sizes: $L=80$
  (data taken from~\cite{janus:09b}), $L=160$ (new simulations) and $L=256$
  (Metropolis dynamics from~\cite{fernandez:15}, rescaling the $x$ axis by a factor of 4
    to compare with our heat-bath dynamics). The
  dashed lines aim to point out the different $\tw$ (and their
  corresponding $\xi$) considered in this work.
  \label{fig:FSSxi}}
\end{figure}

Using heat-bath dynamics on the Janus, Janus II and Memento supercomputers, we
consider the following numerical experiment. Starting from a completely random
configuration of the spins at $T=0.7$, we first let the system evolve in
absence of a magnetic field, i.e. $H=0$, for a waiting time $\tw$. As this $\tw$
grows, the spins rearrange in amorphous magnetic domains of increasing
average size $\xi$, as we show in Fig.~\ref{fig:FSSxi} ($\xi$ is computed with
the $\xi_{12}$ integral estimator described in
Refs.~\cite{janus:08b,janus:09b}). After this time $\tw$, we turn on a tiny
field $H>0$ and follow the response at a later time $t+\tw$. 

We have considered five different values of $\tw$: $t_\mathrm{w}=2^{11}$ and
$t_\mathrm{w}=2^{30}$ were simulated on Janus II; $t_\mathrm{w}=2^{26}, 2^{19}$
and $2^{15}$ on Janus (smaller systems were simulated on Memento, see below our
study of size effects). Times are measured in units of Monte Carlo sweeps. The
measuring times $t$ were chosen as the integer part of $2^{i/4}$ for
integer $i$ (discarding repetitions). For each $\tw$ we repeat the
procedure described above for four values of the magnetic field:
$H\in\{0,0.02,0.04,0.08\}$ in the case of Memento and Janus I supercomputers
and $H\in\{0,0.01,0.02,0.04\}$ on Janus II. We considered exactly the same set
of samples with each $H$ and reused the same sequences of random numbers in an
effort to eliminate sources of fluctuations.

Depending on the computer used, we simulated different system sizes, either
$L=80$ (on Memento and Janus I) or $L=160$ (on Janus II). We simulated 647 samples
for $L=80$ (all $t_\mathrm{w}$ and $H$ values). For $L=160$, we used 55
samples for $t_\mathrm{w}=2^{11}$ and 335 samples for $t_\mathrm{w}=2^{30}$ [we
also simulated 336 samples at $H=0$ in order to compute $\xi(t_\mathrm{w})$].
Notice that self-averaging means that one needs fewer samples for larger sizes.
Previous works at $H=0$ suggested that finite-size effects should be negligible,
compared to our typical statistical accuracy, as long as we ensure
that $L>7\xi(t+t_\mathrm{w})$~\cite{janus:08b}. As a new test of the validity
of this statement, we compare our new results of $\xi(t_\mathrm{w})$ obtained
with Janus II and $L=160$ with previous works corresponding to
$L=80$~\cite{janus:09b} and $L=256$~\cite{fernandez:15} (see
Fig.~\ref{fig:FSSxi}) finding no significant dependence on $L$ in the studied
range of $\tw$.

\section{Computation of the linear susceptibility}\label{sec:susclin}
\begin{figure}[t]
\centering \includegraphics[width=\columnwidth, angle=0]{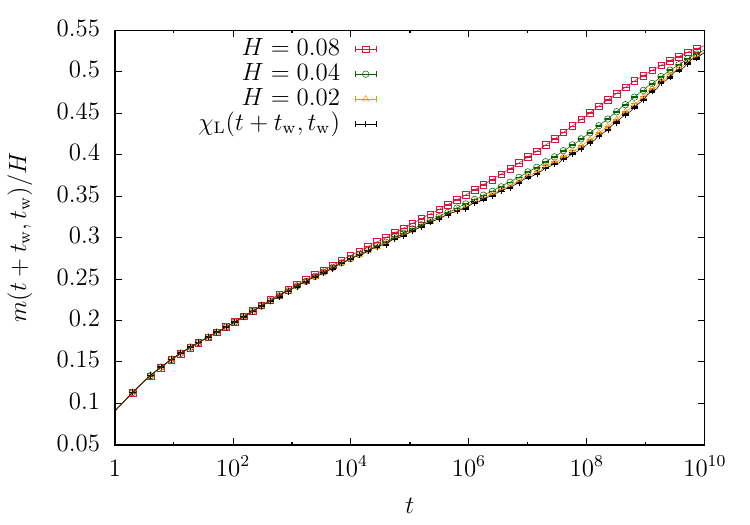}
\caption{ Extraction of the linear susceptibility as a
  function of $t$ from the $m(t+\tw,t_\mathrm{w})/H$ data obtained at
  $H=0.02,\ 0.04$ and $0.08$. Data shown here corresponds to
  $\tw=2^{26}$. For the sake of visibility, only one every two measured times have been plotted in points.
  \label{fig:mh2e26}}
\end{figure}

The discussion on the GFDR requires the computation of
the linear susceptibility, that is, of
\begin{equation}
\chi(t+\tw,t_\mathrm{w})=\left.\frac{\partial m(t+t_\mathrm{w})}{\partial
  H}\right|_{H=0}.
\end{equation}
With this aim, we measure $m(t,t_\mathrm{w})/H$ at several values of
the external field, and use them to extract the $H\to 0$ limit. Indeed,
since the Edwards-Anderson Hamiltonian is odd in the field
around $H=0$, one can write the magnetization in terms of odd powers of $H$, which allows us to separate the linear response $\chi$ from the non-linear responses
\begin{equation}\label{eq:mh}
{m(t+\tw,t_\mathrm{w};H)}=H\chi(t+\tw,t_\mathrm{w})-\frac{H^3}{3!}\chi_\mathrm{NL}(t+\tw,t_\mathrm{w};H).
\end{equation}
In order to make some progress, we Taylor-expand $\chi_\mathrm{NL}=\chi_{3} + \frac{H^2}{20} \chi_{5}+\mathcal{O}(H^4)$, thus finding:
\begin{equation}\label{eq:chiL}
\begin{split}
\frac{m(t+\tw,t_\mathrm{w})}{H}=&\chi(t+\tw,t_\mathrm{w})-\frac{H^2}{3!}\chi_3(t+\tw,t_\mathrm{w})\\
&\quad -\frac{H^4}{5!}\chi_5(t+\tw,t_\mathrm{w})+\mathcal{O}(H^6),
\end{split}
\end{equation}
Therefore, if we measure $m$ for three small fields and neglect higher-order contributions in
$H$, we can extract $\chi(t+\tw,t_\mathrm{w})$ from a set of three
equations and three unknowns
[by the same token, we obtain $\chi_{3}(t+\tw,t_\mathrm{w})$ and
$\chi_5(t+\tw,t_\mathrm{w})$ as well, but these magnitudes will not be discussed herein].
We show in Fig.~\ref{fig:mh2e26} $m(t+\tw,t_\mathrm{w})/H$ and $\chi(t+\tw,t_\mathrm{w})$ for one of our values of $\tw$.

Alternatively, instead of performing simulations at different $H$, one
could have obtained $\chi(t+\tw,t_\mathrm{w})$ directly from
simulations at $H=0$ using methods such as those described in
Refs.~\cite{ricci-tersenghi:03,chatelain:03}. The drawback of this
approach is that it would have required a much larger amount of samples in order
to get equivalent statistical errors.

\section{Smoothing and interpolating the data}\label{sec:smooth}
\begin{figure}[t]
\centering \includegraphics[width=\columnwidth, angle=0]{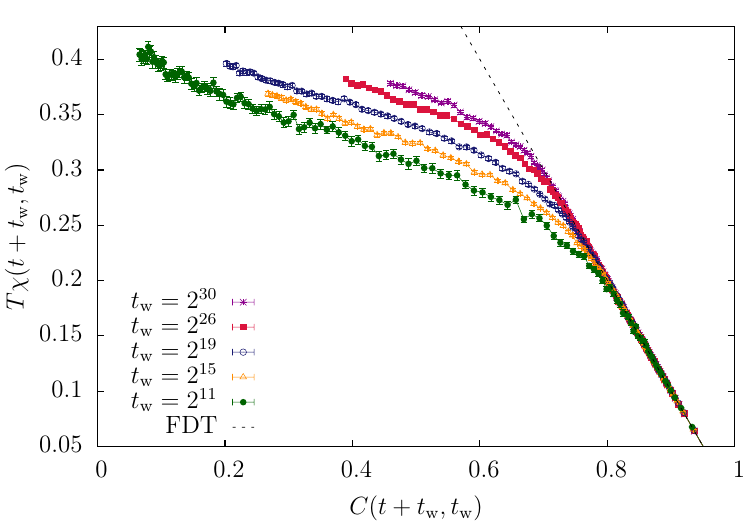}
\caption{ Linear response $T\chi(t,t_\mathrm{w})$
versus $C(t+\tw,t_\mathrm{w})$ at $T=0.7$ and five values of $\tw$ using
raw processed data (to be compared with Fig. 1 in the main text, which was
obtained only after the smoothing of the simulation data at fixed $H$ and an
extrapolation to $H\to 0$). 
  \label{fig:FDToriginal}}
\end{figure}

The original data consisted of pairs 
$\{C(t+\tw,t_\mathrm{w}),\chi(t+\tw,t_\mathrm{w})\}$, where $t$
takes some discrete values. However, if we reproduce Fig.~1 in the main
text but using the raw measurements (see Fig.~\ref{fig:FDToriginal})
we find much noisier curves. Indeed, data for successive times, although very
correlated, displays random fluctuations. Besides, the statistical
errors for $C(t+\tw,t_\mathrm{w};H=0)$ and $C(t+\tw,t_\mathrm{w};H)$ are
completely negligible compared to the errors in
$Tm(t+\tw,t_\mathrm{w};H)/H$ (they are indistinguishable in the figure). We
 used these two facts to our benefit in order to smooth and reduce the
statistical errors of these curves. Let us describe our smoothing procedure
step by step.

\begin{figure}[t]
\centering \includegraphics[width=\columnwidth, angle=0]{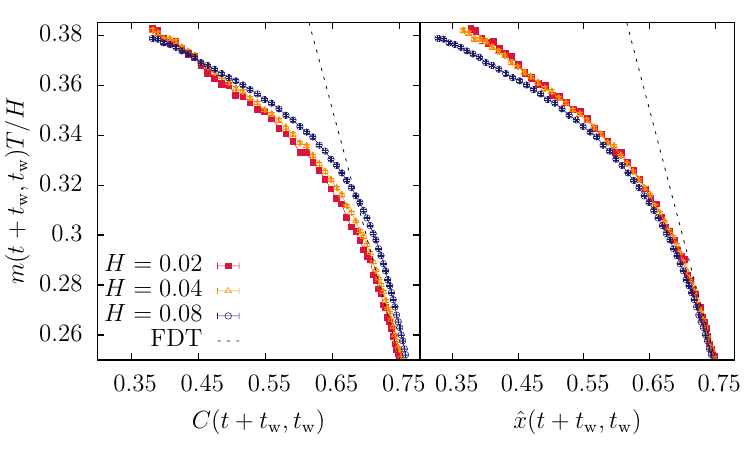}
\caption{ Non-linear corrections in $H$ to  $T\chi(t+\tw,t_\mathrm{w})$ when
plotted versus $C(t+\tw,t_\mathrm{w})$ (left) or $
x(t+\tw,t_\mathrm{w})=C(t+\tw,t_\mathrm{w})+C(t+\tw,t_\mathrm{w};H)/2$ (right). Data
corresponds to $\tw=2^{26}$ and $T=0.7$. \label{fig:FDTdiffC}}
\end{figure}

  We  fit our
data for $Tm(t+\tw,t_\mathrm{w};H)/H$ to a smooth function of
\begin{equation}
\hat x(t+\tw,t_\mathrm{w})=\frac{C(t+\tw,t_\mathrm{w})+C(t+\tw,t_\mathrm{w};H)}{2}\,.
\end{equation}
This choice [instead of just $C(t+\tw,t_\mathrm{w})$], although irrelevant
in the $H\to 0$ limit, turns out to reduce the non-linear corrections
in $H$ as we show in Fig.~\ref{fig:FDTdiffC}, and yields easier and more accurate fits.

Our chosen functional form is as follows. Let the quantity
$Tm(t+\tw,t_\mathrm{w};H)/H$ be approximated by $f(\hat x)$ ($f$
depends on $H$ and $t_\mathrm{w}$, but we will write $f$ nevertheless, to keep
the notation as light as possible):
\begin{equation}\label{eq:fit1}
f(\hat x)=f_\mathrm{L}(\hat x) \frac{1+\tanh[Q(\hat x)] }{2}\,+\, f_\mathrm{S}(\hat x)
\frac{1-\tanh[Q(\hat x)]}{2}\,,
\end{equation}
with $Q(\hat x)=(\hat x-\hat x^*)/w$. In other words, there are two
functional forms: $f_\mathrm{S}$, adequate for small $\hat x$ and $f_\mathrm{L}$,
good for large $\hat x$. The crossover between the two functional
forms takes place at $\hat x^*\approx 0.7$ in an interval of half-width
$w\approx 0.04$ (although we keep $\hat x^*$ and $w$ as fitting
parameters). The functional form for small $\hat x$ are diagonal [$N,N$]
Pad\`e approximants,
\begin{equation}\label{eq:fit2}
f_\mathrm{S}(\hat x)= \frac{\sum_{k=0}^N b_k \hat x^k}{\sum_{k=0}^N a_k \hat x^k}\,.
\end{equation}
As for the region where deviations from the fluctuation-dissipation theorem
are tiny, we chose a polynomial in $1-\hat x$
\begin{equation}\label{eq:fit3}
f_\mathrm{L}(\hat x)= (1-\hat x) + \sum_{k=2}^{N'} c_k  (1-\hat x)^k.
\end{equation}
We keep ${a_k,b_k,c_k}$ as fitting variables.

Following Refs.~\cite{janus:08b,janus:09b,fernandez:15,lulli:15}, we
perform a fit considering only the diagonal part of the covariance
matrix (we obtain $\chi^2/\mathrm{DOF}$ significantly smaller than one,
probably due to data correlation). Errors are computed following a
jackknife procedure [we perform an independent fit for each jackknife
block, and compute errors from the jackknife fluctuations of the
fitted $f(\hat x)$]. Our fits are reported in Table~\ref{table:fits}.

\begin{table}
\centering
\caption{Information about the fits to Eqs.~(\ref{eq:fit1},\ref{eq:fit2},\ref{eq:fit3}).\label{table:fits}}
\begin{ruledtabular}
\begin{tabular}{ccccr}
\boldmath $t_\mathrm{w}$ & \boldmath $H$ &  \boldmath $N$ &  \boldmath $N'$ &  \boldmath $\chi^2/\mathrm{DOF}$\\
\hline
 & 0.01 & 2 & 1 & 51.6822/127 \\
 $2^{11}$       & 0.02 & 2 & 1 & 43.9926/127 \\
        & 0.04 & 2 & 1 & 45.6321/127 \\
\hline
 & 0.02 & 2 & 1 & 33.1259/90 \\
 $2^{15}$       & 0.04 & 2 & 1 & 43.3823/90  \\
        & 0.08 & 2 & 2 & 21.0832/89  \\
\hline
& 0.02 & 3 & 2 & 27.6364/115    \\
 $2^{19}$        & 0.04 & 3 & 2 &  25.8737/115   \\
        & 0.08 & 3 & 3 &  31.6819/114    \\
\hline
& 0.02 & 2 & 1 &  29.5259/118      \\
 $2^{26}$        & 0.04 & 2 & 1 &  36.5544/118   \\
        & 0.08 & 2 & 1 &  57.3693/118      \\
\hline
& 0.01 & 2 & 1 & 31.7369/126  \\
 $2^{30}$        & 0.02 & 3 & 3 &  24.7701/122 \\
        & 0.04 & 3 & 2 &  33.0019/123 \\
\end{tabular}
\end{ruledtabular}
\centering
\end{table}

\begin{figure}[t]
\centering \includegraphics[width=\columnwidth, angle=0]{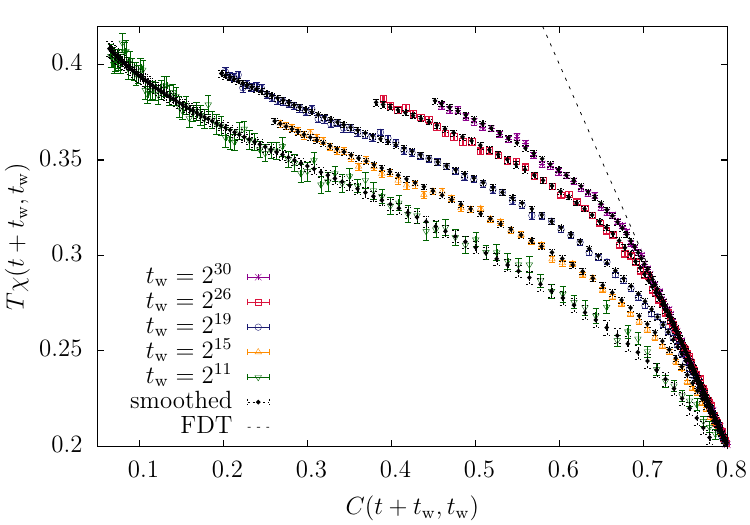}
\caption{ Comparison between the original (in color
  and empty dots) and smoothed data (in black full dots) in the
  Linear response $T\chi(t+\tw,t_\mathrm{w})$ versus
  $C(t+\tw,t_\mathrm{w})$ curves. Data corresponds to $T=0.7$ and five
  values of $\tw$.  \label{fig:FDToriginal-smoothed}}
\end{figure}

\begin{figure}[t]
\centering \includegraphics[width=\columnwidth, angle=0]{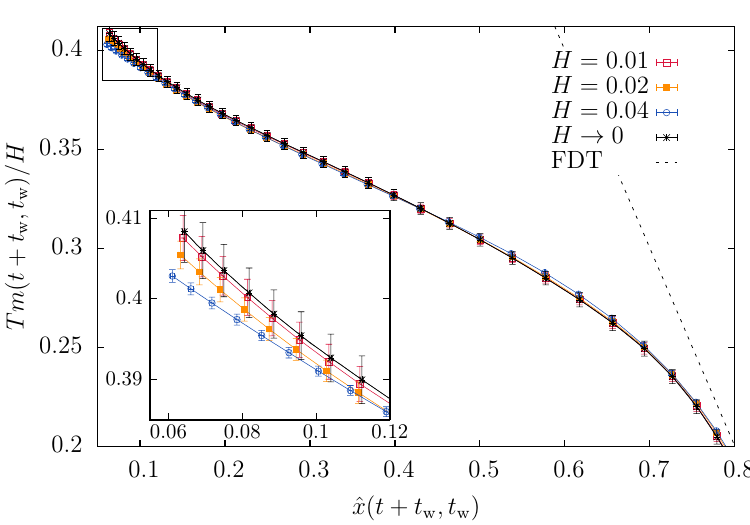}
\caption{ $Tm(t+\tw,t_\mathrm{w})/H$ versus $\hat{x}(t+\tw,t_\mathrm{w})$ for several values of $H$ (in color empty dots) and $\tw=2^{11}$, together with the extrapolation $H\to 0$ (in black crosses). The inset is a blow up of the region for large $t/\tw$ in the square box. 
  \label{fig:2e11}}
\end{figure}

Once each curve $Tm(t+\tw,t_\mathrm{w})/H$ is smoothed at each $H$, we
extract the linear susceptibility following the procedure described in
the previous Section. We show a comparison between the original and
smoothed data in Fig.~\ref{fig:FDToriginal-smoothed}. We found that in
most the cases the extrapolated linear response
$T\chi(t+\tw,t_\mathrm{w})$ was compatible within the error
with the smaller field considered. However, the
extrapolation $H\to 0$ becomes particularly delicate and even changes the
shape of the curve at large values of the $t/\tw$ ratio, as we show
in Fig.~\ref{fig:2e11}.

\section[Fit of $S(C,L)$ and computation of $L_\mathrm{eff}$]{\boldmath Fit of $S(C,L)$ and computation of $L_\mathrm{eff}$}\label{sec:Leff}
\begin{figure}[t]
\centering \includegraphics[width=\columnwidth, angle=0]{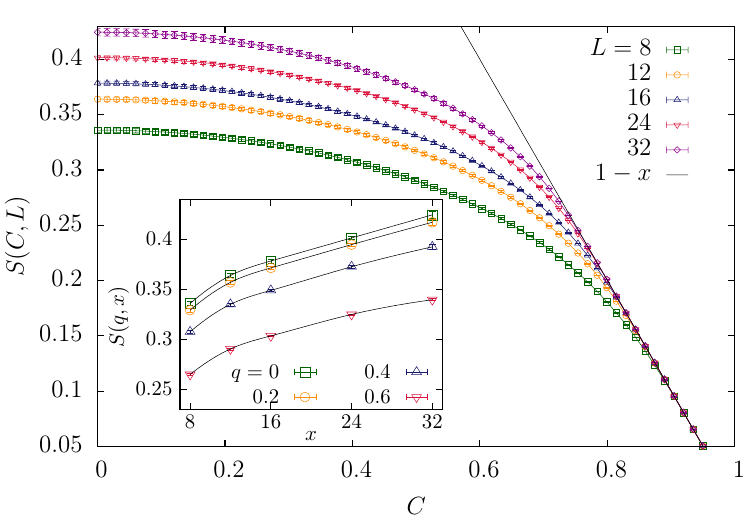}
\caption{ $S(C,L)$ versus $C$ for different system sizes
  obtained using~\eqref{eq:SCL} and data from
  Ref.~\cite{janus:10}. (Inset) Orthogonal cuts to the figure in the
  main panel plotted as function of $L$ in color points together with
  the interpolating cubic spline curve along this variable.\label{fig:SCL}}
\end{figure}

Part of our discussion in the main text seeks to find a relation
between the linear response at finite $\tw$ with the overlap distribution
$P(q,L)$ in equilibrium at a finite size $L_{\mathrm{eff}}$.
That is,
\begin{equation}\label{eq:FDTV-2-app}
T\chi(t+\tw,t_\mathrm{w})= S\big(C(t+\tw,t_\mathrm{w}),L_{\mathrm{eff}}(t+\tw, 
t_\mathrm{w})\big)\,,
\end{equation}
where 
\begin{equation}
S(C,L)=\int_C^1 d\mathrm{C'}\, x(C',L)\,,\  x(C,L)=\int_{0}^C\,\mathrm{d} q\, 2 
P(q,L)\,.
\end{equation}
We computed $S(C,L)$ by means of a numerical integration
of the  $P(q,L)$ discussed in Ref.~\cite{janus:10} for
$L=8,\ 12,\ 16,\ 24$ and $32$. We show  $S(C,L)$ in the main panel of
Fig.~\ref{fig:SCL}. In order to
identify $L_{\mathrm{eff}}$ we needed a function $S(q,x)$
that is continuous both in $C$ and in $L$, which we construct by 
computing a cubic
spline\footnote{We do not used the so-called ``natural'' cubic spline. Instead, we fixed
  the first and last derivative of the interpolating function from three
  points of a parabolic fit.} of the data along both variables (first
in $C$ and only then in $L$). Errors are computed using the jackknife method.
We show some interpolation curves along the
$x$ variable in the inset of Fig.~\ref{fig:SCL}. Once $S(q,x)$ is at
hand, $L_{\mathrm{eff}}(t+t_\mathrm{w}, t_\mathrm{w})$ can be
extracted by looking for the $x$ value that satisfies~\eqref{eq:FDTV-2-app} at each time $t$, fixing the off-equilibrium data
$T\chi(t+\tw,t_\mathrm{w})$ and $C(t,t_\mathrm{w})$.

\section{Finite-size effects in the response}\label{sec:finite-size}

\begin{figure}[t]
\centering \includegraphics[width=\columnwidth, angle=0]{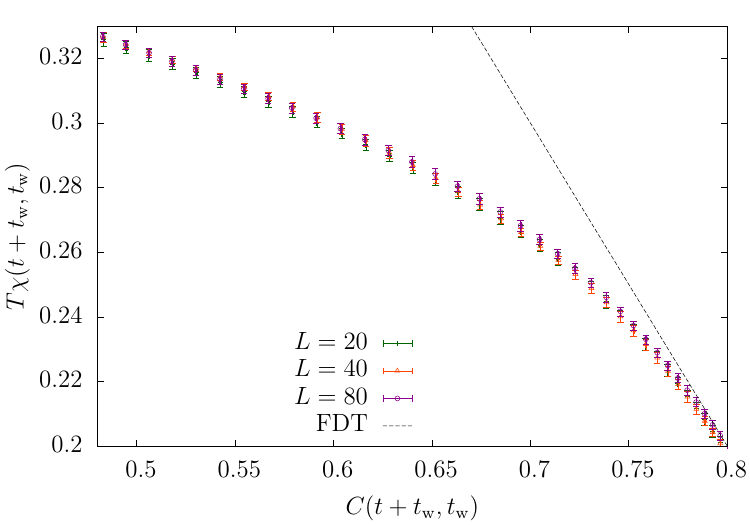}
\caption{ Absence of finite-size effects in the response
  function $T \chi(t+\tw,t_\mathrm{w})$ versus $C(t+\tw,t_\mathrm{w})$ at
  $T=0.7$. Data from $L=20,\ 40$ and $80$ are compared in the case of
  $t_\mathrm{w}=2^{15}$. All the points are compatible within the error bars.\label{fig:FSS}}
\end{figure}

Up to now, finite-size effects have been investigated only for single-time
correlation functions [and the related extraction of
$\xi(t_\mathrm{w})$]. As far as we know, size effects were not studied
previously in the response to a magnetic field
$\chi(t+\tw,t_\mathrm{w})$. In this context, it is somewhat worrying that
we have identified a large length scale $L_\mathrm{eff}\approx 100$
(discussed below) in the regime where deviations from the FDT are incipient.
For this reason, we have explicitly checked
that our data does not suffer from finite-size effects in that region (as
we show in Fig.~\ref{fig:FSS}) by comparing results from three system
sizes, $L=20,\ 40$ and $80$, in the case of $t_\mathrm{w}=2^{15}$,
finding no finite-size dependence. For the smaller system sizes we
considered 28000 samples for $L=20$ and 12000 samples for $L=40$.

\section{A simple inequality}\label{sec:inequality}

In the main text, we have used several times the inequality
\begin{equation}\label{eq:desigualdad}
S(C,L) \leq 1 -
\overline{\langle |q| \rangle }_{L=\infty}\,. 
\end{equation}
Our purpose here is to remind the reader of its derivation, for the
sake of completeness.

Let us first recall the notations used in the main text:
\begin{eqnarray}\label{eq:SCL1}
S(C,L)&=&\int_C^1 d\mathrm{C'}\, x(C',L)\,,\\  
x(C,L)&=&\int_{0}^C\,\mathrm{d} q\, 2 P(q,L)\,.\label{eq:SCL2}
\end{eqnarray}
We start by noticing
\begin{equation}
S(C,L) \leq S(C=0,L)\,,
\end{equation}
due to the inequality $x(C,L)\geq 0$ for the cumulative distribution.
Next, we integrate by parts to find [recall that $P(q,L)=P(-q,L)$]
\begin{equation}
S(C=0,L)=1-\overline{\langle |q| \rangle }_L\,,\quad \overline{\langle |q| \rangle }_L\equiv \int_{-1}^1\, \mathrm{d} q |q| P(q,L)\,.
\end{equation}
Finally, to obtain the upper bound in~(\ref{eq:desigualdad}), we remark that $\overline{\langle |q| \rangle }_L$ is
monotonically decreasing in $L$ for a system with periodic boundary
conditions.

\section{The ferromagnetic case and conditions for validity of Eq.~(5) of main text}\label{sec:ferromagnet}

\begin{figure}[t]
\centering \includegraphics[width=\columnwidth, angle=0]{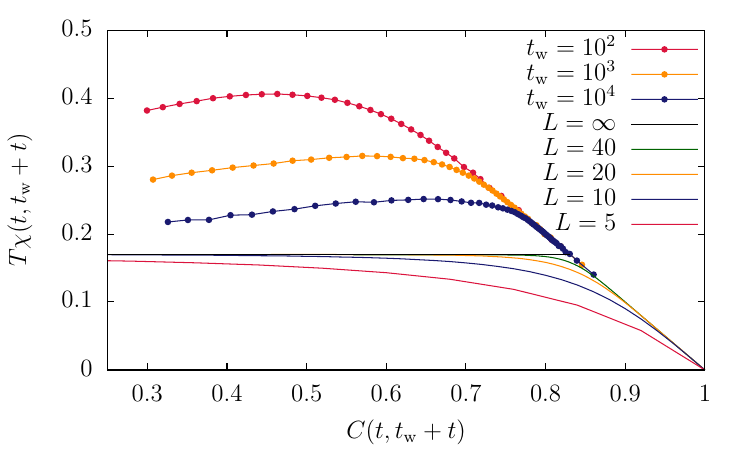}
\caption{Upper lines with points are data for $T \chi(t+\tw,t_\mathrm{w})$ versus $C(t+\tw,t_\mathrm{w})$ measured in the $D=2$ ferromagnetic Ising model at $T=2\approx 0.88 T_c$ (data from Ref.~\cite{ricci-tersenghi:03}). Lower lines are the equilibrium $S(C,L)$ for the same model and their thermodynamic limit $S(C,L=\infty)$.
For this model Eq.~(5) of main text can not be satisfied and the SDD does not exist. \label{fig:SCferro}}
\end{figure}

Our SDD is based on Eq.~(5) in the main text that we repeat here for readers convenience
\begin{equation}
T\chi(t+\tw,t_\mathrm{w}) =
S\big(C(t+\tw,t_\mathrm{w}),L_{\mathrm{eff}}(t+t_\mathrm{w}, t_\mathrm{w})\big)\,.
\label{eq:Leff}
\end{equation}
Although for the $D=3$ Edwards-Anderson (EA) model the above equation can be satisfied for all our data, it is not obvious that this is the case for other models.
In particular we show in Fig.~\ref{fig:SCferro} a simple case where \eqref{eq:Leff} can not be satisfied.

In Fig.~\ref{fig:SCferro} we show both equilibrium and non-equilibrium data for the $D=2$ ferromagnetic Ising model gathered at temperature $T=2\approx 0.88 T_c$.
For the non-equilibrium data we reproduce correlation and responses already published in Ref.~\cite{ricci-tersenghi:03}, while the equilibrium data have been obtained by running the Wolff algorithm \cite{wolff:89}.
The black line is the thermodynamical limit for the equilibrium data
\[
S(C,\infty) = \min(1-m(T)^2, 1-C)
\]
where $m(T)$ is the remanent magnetization.

It is clear from data in Fig.~\ref{fig:SCferro} that there is no $L_{\mathrm{eff}}$ size such that the non-equilibrium data can be matched with the equilibrium ones. This is a direct consequence of the fact that finite size effects in this model are such that $S(C,L) \le S(C,\infty)$, while the dynamical curves show an excess of response, bringing them above $S(C,\infty)$.

In general the condition for the applicability of \eqref{eq:Leff} is that the dynamical curves must lie in the region of the $(T \chi, C)$ plane covered by the equilibrium functions $S(C,L)$. In the present case such a region is very narrow (as shown in Fig.~\ref{fig:SCferro} for $L \ge 5$) and the dynamical curves miss it. Luckily enough the analogous region for the $D=3$ EA model is very wide, and \eqref{eq:Leff} can be always satisfied on the timescales we have probed.

The very different behaviour between the above two models can be explained by noticing that there are at least two major sources of finite times effects:
\begin{itemize}
\item the first is the one discussed thoroughly in the main text. Its application to the ferromagnetic Ising model should give a really tiny effect, because the $S(C,L)$ converges very fast to its thermodynamical limit;
\item the second correction comes from the convergence of one-time quantities (e.g. the energy density) to their large time limit. This is the dominating one for the ferromagnetic Ising model, where the energy density decays as $E(t) - E(\infty) \propto \xi(t)^{-b}$, with $b=1$. We expect this contribution to be much less important in the EA model, since the exponent is $b \simeq 2.6$ \cite{janus:09b}.
The ferromagnetic Ising model is very peculiar; in the general case, using the hand-waiving argument that the exponent $b$ equals the lower critical dimension, we expect $b>1$ (e.g.\ $b=2$ in models with continuous variables) and this correction to be much less relevant.
\end{itemize}

\section{Extrapolating the effective size}\label{sec:Pdyn}
We have shown in the main text that, for every $t_\mathrm{w}$ and
small enough $t$, $L_\mathrm{eff}(t+\tw,t_\mathrm{w})$ can be very
large. This \emph{short-time but large-size} effect arises when
$C(t+\tw,t\mathrm{w})\approx q_\mathrm{EA}^{L=4\xi(t_\mathrm{w})}$. In
fact, for $t_\mathrm{w}=2^{30}$ (our largest) 
we can compute $L_\mathrm{eff}$ without extrapolations only for the largest $t$.

The above observation begs the question: how large can
$L_\mathrm{eff}$ be in this small-$t$ regime? We provide here a crude
extrapolation for our $t_\mathrm{w}=2^{30}$ data, mostly based on the
scaling laws found in~\cite{janus:10}.

We start by noticing that one could be tempted to extract the
spin-overlap probability directly from the aging response. One can define
the  \emph{dynamic overlap} probability density function:
\begin{equation}\label{eq:p_dyn_q}
P_\mathrm{dyn}(q;t_\mathrm{w})=-\frac{1}{2}\left.\frac{\partial^2
T\chi(C,t_\mathrm{w})}{\partial C^2}\right|_{C=q}\,.
\end{equation}
Then, one could compare $P_\mathrm{dyn}$ with the equilibrium $P(q,L)$
at $q=C(t+\tw,t_\mathrm{w})$. 
The weak point in this approach is that taking two derivatives of the
curve $T\chi(C,t_\mathrm{w})$, which is subject to random
errors, is very difficult. 

Our way out will be to recall that the area under the peak of the
$P(q,L)$ is approximately $L$-independent~\cite{janus:10}. Therefore,
we shall estimate the peak height (rather than the
peak width).

\begin{figure}[t]
\centering \includegraphics[width=\columnwidth]{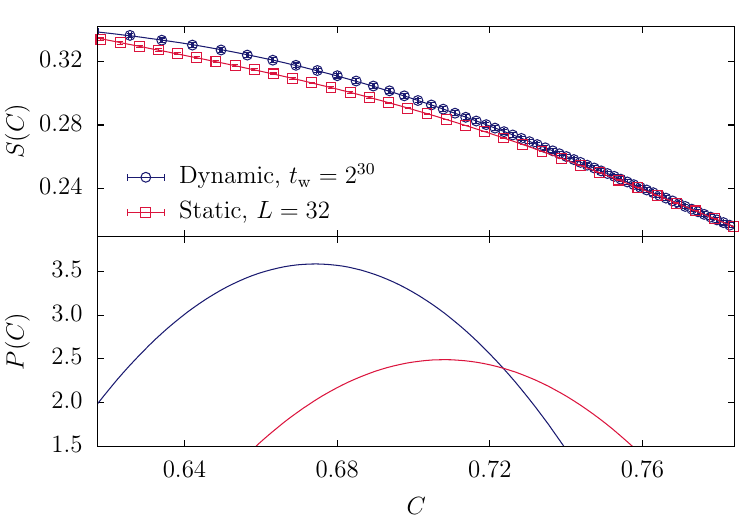}
\caption{ {\bf Numerical attempt to locate the maximum of
    $P_{\mathrm{dyn}}(q,t_{\mathrm{w}}=2^{30})$.} In the top panel, we
  compare the dynamic response $T\chi(C,t_\mathrm{w}=2^{30})$ with the
  equilibrium curve $S(C,L=32)$. The range of $C$ covers the peak
  width of $P(q,L=32)$~\cite{janus:10}.  Since the curvature is
  clearly larger for $T\chi$ than for $S(C,L=32)$, \eqref{eq:p_dyn_q}
  tells us that that the maximum of
  $P_\mathrm{dyn}(q;t_\mathrm{w}=2^{30})$ is higher than the maximum
  of $P(q,L=32)$. The lines correspond to diagonal fits to fourth
  order polynomials in $C$ (we increased the order of the polynomial
  until the figure of merit diagonal-$\chi^2$ for the fit of the
  dynamic response no longer decreased). The bottom panel
  shows the second derivative of the interpolating polynomials of the
  top panel, multiplied by $-1/2$. According to ~\eqref{eq:p_dyn_q},
  these derivatives should give us $P_{\mathrm{dyn}}(q,t_\mathrm{w})$
  and $P(q,L)$. Indeed, the peak position and height in $P(q,L=32)$ is
  very reasonably reproduced by this approach, see
  Ref.~\cite{janus:10}.}\label{fig:P-dyn}
\end{figure}

Our efforts to locate the maximum (let alone the full curve) for
$P_\mathrm{dyn}(q;t_\mathrm{w}=2^{30})$ are documented in
Fig.~\ref{fig:P-dyn} (but the reader is warned to take the results
\emph{cum grano salis}). We note from Fig.~\ref{fig:P-dyn} that the
ratio of the height of the maxima for $t_\mathrm{w}=2^{30}$ and $L=32$ is
$\sim 3.6/2.5$.  Therefore, from the scaling of the peak width,
$\propto L^{-B\approx 0.28}$, we extrapolate
\begin{equation}
L_\mathrm{eff}\sim 32 \times (3.6/2.5)^{\frac{1}{B}}\approx 118\,,
\end{equation}
which is certainly larger than our maximum equilibrium size, $L=32$.

\section[The simplified $S(C,L)$]{\boldmath The simplified $S(C,L)$}

In the main text, we wondered about the consequences of having
at our disposal only a simplified approximation for $S(C,L)$:
\begin{equation}
S_\mathrm{simpl}(C,L)=\min\left[S_0(L)- P_0 C^2-\frac{P_1}{6} C^4, 1-C\right]\,.
\end{equation}
In the above equation, $P_0$ and $P_1$ are $L$-independent
constants. All the depedence on the system size is in $S_0(L)$. In
fact, $S_0(L)$ was obtained by fitting the actual data
$S(C=0,L=8,12,16,24,32)$ to a quadratic polynomial in $L^{-\theta}$. We
took $\theta=0.38$ from Ref.~\cite{janus:10} [recall that the maximum
of the spin-overlap probability, $P(q,L)$ scales with $L$ as
$q_\mathrm{EA}^{(L)}-q_\mathrm{EA}^{(\infty)}\propto
L^{-\theta}$]. Once $S_0(L)$ was known, we determined the
constants $P_0$ and $P_1$ from a least-squares minimization of the
difference between $S_\mathrm{simpl}(C,L)$ and the actual data.

\end{document}